\documentclass[twocolumn,aps,prx,showpacs,preprintnumbers,amsmath,amssymb, superscriptaddress]{revtex4-1}
\usepackage{bm}
\usepackage{graphicx}
\usepackage{xcolor,xspace}
\usepackage{ulem}
\usepackage{float}
\usepackage[colorlinks=true,
			linkcolor=blue,
			urlcolor=blue,
			citecolor=blue]{hyperref}
\usepackage{hyperref}
\usepackage{geometry}
\begin{document}
\title{Evidence for an Excitonic Insulator State in $\bf{Ta_{2}Pd_{3}Te_5}$}

\author{Jierui Huang}
\thanks{These authors contributed equally to this work.}
\affiliation{Beijing National Laboratory for Condensed Matter Physics and Institute of
Physics, Chinese Academy of Sciences, Beijing 100190, China}
\affiliation{University of Chinese Academy of Sciences, Beijing 100049, China}
\author{Bei Jiang}
\thanks{These authors contributed equally to this work.}
\affiliation{Beijing National Laboratory for Condensed Matter Physics and Institute of
Physics, Chinese Academy of Sciences, Beijing 100190, China}
\affiliation{University of Chinese Academy of Sciences, Beijing 100049, China}
\author{Jingyu Yao}
\thanks{These authors contributed equally to this work.}
\affiliation{Beijing National Laboratory for Condensed Matter Physics and Institute of
Physics, Chinese Academy of Sciences, Beijing 100190, China}
\affiliation{University of Chinese Academy of Sciences, Beijing 100049, China}
\author{Dayu Yan}
\thanks{These authors contributed equally to this work.}
\affiliation{Beijing National Laboratory for Condensed Matter Physics and Institute of
Physics, Chinese Academy of Sciences, Beijing 100190, China}
\affiliation{University of Chinese Academy of Sciences, Beijing 100049, China}
\author{Xincheng Lei}
\thanks{These authors contributed equally to this work.}
\affiliation{Beijing National Laboratory for Condensed Matter Physics and Institute of
Physics, Chinese Academy of Sciences, Beijing 100190, China}
\affiliation{University of Chinese Academy of Sciences, Beijing 100049, China}
\author{Jiacheng Gao}
\affiliation{Beijing National Laboratory for Condensed Matter Physics and Institute of
Physics, Chinese Academy of Sciences, Beijing 100190, China}
\affiliation{University of Chinese Academy of Sciences, Beijing 100049, China}
\author{Zhaopeng Guo}
\affiliation{Beijing National Laboratory for Condensed Matter Physics and Institute of
Physics, Chinese Academy of Sciences, Beijing 100190, China}
\author{Feng Jin}
\affiliation{Beijing National Laboratory for Condensed Matter Physics and Institute of
Physics, Chinese Academy of Sciences, Beijing 100190, China}
\author{Yupeng Li}
\affiliation{Beijing National Laboratory for Condensed Matter Physics and Institute of
Physics, Chinese Academy of Sciences, Beijing 100190, China}
\author{Zhenyu Yuan}
\affiliation{Beijing National Laboratory for Condensed Matter Physics and Institute of
Physics, Chinese Academy of Sciences, Beijing 100190, China}
\affiliation{University of Chinese Academy of Sciences, Beijing 100049, China}
\author{Congcong Chai}
\affiliation{Beijing National Laboratory for Condensed Matter Physics and Institute of
Physics, Chinese Academy of Sciences, Beijing 100190, China}
\affiliation{University of Chinese Academy of Sciences, Beijing 100049, China}
\author{Haohao Sheng}
\affiliation{Beijing National Laboratory for Condensed Matter Physics and Institute of
Physics, Chinese Academy of Sciences, Beijing 100190, China}
\affiliation{University of Chinese Academy of Sciences, Beijing 100049, China}
\author{Mojun Pan}
\affiliation{Beijing National Laboratory for Condensed Matter Physics and Institute of
Physics, Chinese Academy of Sciences, Beijing 100190, China}
\affiliation{University of Chinese Academy of Sciences, Beijing 100049, China}
\author{Famin Chen}
\affiliation{Beijing National Laboratory for Condensed Matter Physics and Institute of
Physics, Chinese Academy of Sciences, Beijing 100190, China}
\affiliation{University of Chinese Academy of Sciences, Beijing 100049, China}
\author{Junde Liu}
\affiliation{Beijing National Laboratory for Condensed Matter Physics and Institute of
Physics, Chinese Academy of Sciences, Beijing 100190, China}
\affiliation{University of Chinese Academy of Sciences, Beijing 100049, China}
\author{Shunye Gao}
\affiliation{Beijing National Laboratory for Condensed Matter Physics and Institute of
Physics, Chinese Academy of Sciences, Beijing 100190, China}
\affiliation{University of Chinese Academy of Sciences, Beijing 100049, China}
\author{Gexing Qu}
\affiliation{Beijing National Laboratory for Condensed Matter Physics and Institute of
Physics, Chinese Academy of Sciences, Beijing 100190, China}
\affiliation{University of Chinese Academy of Sciences, Beijing 100049, China}
\author{Bo Liu}
\affiliation{Beijing National Laboratory for Condensed Matter Physics and Institute of
Physics, Chinese Academy of Sciences, Beijing 100190, China}
\affiliation{University of Chinese Academy of Sciences, Beijing 100049, China}
\author{Zhicheng Jiang}
\affiliation{Shanghai Synchrotron Radiation Facility, Shanghai Advanced Research Institute,
Chinese Academy of Sciences, Shanghai 201204, China}
\author{Zhengtai Liu}
\affiliation{Shanghai Synchrotron Radiation Facility, Shanghai Advanced Research Institute,
Chinese Academy of Sciences, Shanghai 201204, China}
\author{Xiaoyan Ma}
\affiliation{Beijing National Laboratory for Condensed Matter Physics and Institute of
Physics, Chinese Academy of Sciences, Beijing 100190, China}
\affiliation{University of Chinese Academy of Sciences, Beijing 100049, China}
\author{Shiming Zhou}
\affiliation{Hefei National Research Center for Physical Sciences at the Microscale,
University of Science and Technology of China, Hefei 230026, China}
\author{Yaobo Huang}
\affiliation{Shanghai Synchrotron Radiation Facility, Shanghai Advanced Research Institute,
Chinese Academy of Sciences, Shanghai 201204, China}
\author{Chenxia Yun}
\affiliation{Beijing National Laboratory for Condensed Matter Physics and Institute of
Physics, Chinese Academy of Sciences, Beijing 100190, China}
\author{Qingming Zhang}
\affiliation{Beijing National Laboratory for Condensed Matter Physics and Institute of
Physics, Chinese Academy of Sciences, Beijing 100190, China}
\affiliation{School of Physical Science and Technology, Lanzhou University, Lanzhou 730000, China}
\author{Shiliang Li}
\affiliation{Beijing National Laboratory for Condensed Matter Physics and Institute of
Physics, Chinese Academy of Sciences, Beijing 100190, China}
\affiliation{University of Chinese Academy of Sciences, Beijing 100049, China}
\affiliation{Songshan Lake Materials Laboratory, Dongguan 523808, China}
\author{Shifeng Jin}
\affiliation{Beijing National Laboratory for Condensed Matter Physics and Institute of
Physics, Chinese Academy of Sciences, Beijing 100190, China}
\affiliation{University of Chinese Academy of Sciences, Beijing 100049, China}
\author{\\Hong Ding}
\affiliation{Tsung-Dao Lee Institute, New Cornerstone Science Laboratory, and School of Physics and Astronomy, Shanghai Jiao Tong University, Shanghai 201210, China}
\affiliation{Hefei National Laboratory, Hefei 230088, China}
\author{Jie Shen}
\email[Corresponding author.]{shenjie@iphy.ac.cn}
\affiliation{Beijing National Laboratory for Condensed Matter Physics and Institute of
Physics, Chinese Academy of Sciences, Beijing 100190, China}
\author{Dong Su}
\email[Corresponding author.]{dongsu@iphy.ac.cn}
\affiliation{Beijing National Laboratory for Condensed Matter Physics and Institute of
Physics, Chinese Academy of Sciences, Beijing 100190, China}
\affiliation{University of Chinese Academy of Sciences, Beijing 100049, China}
\author{Youguo Shi}
\email[Corresponding author.]{ygshi@iphy.ac.cn}
\affiliation{Beijing National Laboratory for Condensed Matter Physics and Institute of
Physics, Chinese Academy of Sciences, Beijing 100190, China}
\affiliation{University of Chinese Academy of Sciences, Beijing 100049, China}
\affiliation{Songshan Lake Materials Laboratory, Dongguan 523808, China}
\author{Zhijun Wang}
\email[Corresponding author.]{wzj@iphy.ac.cn}
\affiliation{Beijing National Laboratory for Condensed Matter Physics and Institute of
Physics, Chinese Academy of Sciences, Beijing 100190, China}
\affiliation{University of Chinese Academy of Sciences, Beijing 100049, China}
\author{Tian Qian}
\email[Corresponding author.]{tqian@iphy.ac.cn}
\affiliation{Beijing National Laboratory for Condensed Matter Physics and Institute of
Physics, Chinese Academy of Sciences, Beijing 100190, China}

\begin{abstract}
The excitonic insulator (EI) is an exotic ground state of narrow-gap semiconductors and semimetals, arising from spontaneous condensation of electron-hole pairs bound by attractive Coulomb interaction. Despite research on EIs dating back to half a century ago, their existence in real materials remains a subject of ongoing debate. In this study, through systematic experimental and theoretical investigations, we have provided evidence for the existence of an EI ground state in a van der Waals compound $\rm{Ta_2Pd_3Te_5}$. Density functional theory calculations suggest that it is a semimetal with a small band overlap, whereas various experiments exhibit an insulating ground state with a clear band gap. Upon incorporating electron-hole Coulomb interaction into our calculations, we obtain an EI phase where the electronic symmetry breaking opens a many-body gap. Angle-resolved photoemission spectroscopy measurements exhibit that the band gap is closed with a significant change in the dispersions as the number of thermally excited charge carriers becomes sufficiently large in both equilibrium and non-equilibrium states. Structural measurements reveal a slight breaking of crystal symmetry with exceptionally small lattice distortion in the insulating state, which cannot account for the significant gap opening. Therefore, we attribute the insulating ground state with a gap opening in $\rm{Ta_2Pd_3Te_5}$ to exciton condensation, where the coupling to the symmetry-breaking electronic state induces a subtle change in the crystal structure.
\end{abstract}


\maketitle
\section{Introduction}\label{section1}
		
In condensed matter systems, many-body interactions can lead to various exotic quantum phases, such as the Mott insulator, unconventional superconductivity, quantum spin liquid, heavy fermion, and so on. One particularly intriguing phenomenon is the excitonic insulator (EI), expected to emerge when a semiconductor with a small band gap or a semimetal with a small band overlap is cooled to sufficiently low temperatures \cite{Mott1961,Halperin1968,Kohn1967,Jerome1967}. During this process, bound electron-hole pairs, known as excitons, spontaneously form and condensate into a phase-coherent insulating state. While exciton condensation has been achieved in artificial heterostructures \cite{Butov1994,Butov1998,Datta1985,Xia1992,Naveh1996,Du2017,Min2008,Li2017,Wang2019,Jauregui2019,Ma2021}, the existence of an EI state in real materials remains a topic of ongoing debate.

So far, only a few materials have been proposed as candidates for EIs. Among them, the most intensively studied ones are 1\textit{T}-$\rm{TiSe_2}$ \cite{Wilson1977,Wilson1978,Pillo2000,Kidd2002,Cercellier2007,Monney2010,Kogar2017} and $\rm{Ta_2NiSe_5}$ \cite{Wakisaka2009,Kaneko2013,Seki2014,Larkin2017,Lu2017,Sugimoto2018,Fukutani2021}, in which a band gap is opened around the Fermi level ($\textit{E}\rm{_F}$) below a critical temperature \cite{Monney2010,Seki2014}. The gap opening is accompanied by significant structural changes, characterized by a charge-density-wave transition with periodic lattice distortion in 1\textit{T}-$\rm{TiSe_2}$ \cite{Salvo1976,Woo1976} and an orthorhombic-to-monoclinic transition with mirror symmetry breaking in $\rm{Ta_2NiSe_5}$ \cite{Sunshine1985,Salvo1986}. The coexistence of electronic and structural instabilities has aroused intense debate on whether the gap opening originates from exciton condensation or lattice distortion. In the former case, excitons spontaneously form and condense due to electron-hole coupling, leading to the opening of a many-body gap. In the EI state, the electronic symmetry breaking induces lattice distortion through electron-lattice coupling \cite{Pillo2000,Cercellier2007,Kaneko2013,Mazza2020}. In the latter case, a structure phase transition is driven by phononic instability. The lattice distortion couples to electrons, opening a hybridized gap. Density functional theory (DFT) calculations for these two materials have revealed the presence of imaginary phonon frequencies, indicating an unstable crystal structure \cite{Calandra2011,Duong2015,Otto2021,Windgatter2021,Subedi2020,Baldini2020}. Despite numerous efforts over several decades, the origin of the gap opening remains a subject of significant controversy.

\begin{figure*}[t]
\centering
\includegraphics[width=0.7\textwidth]{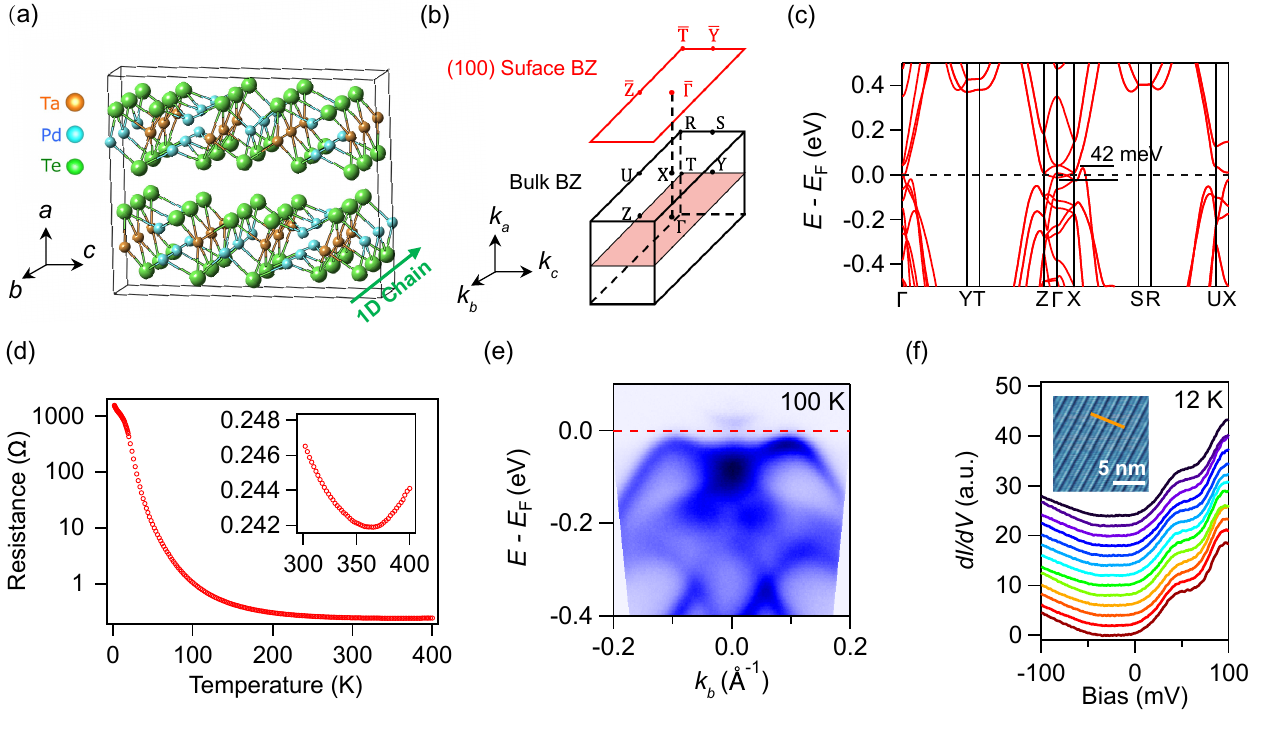}
 \caption{(a) Crystal structure of $\rm{Ta_2Pd_3Te_5}$. (b) Bulk Brillouin zone (BZ)  and (100) surface BZ. (c) Calculated band structure along high-symmetry lines by DFT using the MBJ functional. (d) Resistance as a function of temperature ($R \sim T$) in the $b - c$ plane. The inset shows a metal-insulator transition around 365 K. (e) Intensity plot of the ARPES data along $\bar{\Gamma} - \bar{Y}$ collected with $h\nu$ = 6 eV at 100 K. For clarity, the data are divided by the Fermi-Dirac distribution function to visualize the VB bottom above $E_{\rm{F}}$. (f) Tunneling differential conductance spectra at 12 K along the yellow line in the inset. The inset shows a constant current topographic image of the cleaved (100) surface ($V$ = 100 mV, $I$ = 100 pA). The data were collected in a clean region free from impurities to display the global gap.}
\label{Fig. 1}
 \end{figure*}
    
As the symmetry of the electronic state is broken in an EI state, structural instability inevitably occurs due to electron-lattice coupling, which is inherently present in solid-state materials. Nevertheless, as a by-product of exciton condensation, lattice distortion can, in principle, be significantly suppressed. If the effects of structural changes on the band structure become sufficiently small, the structural origin of the gap opening can be ruled out. In this work, we have shown that the van der Waals compound $\rm{Ta_2Pd_3Te_5}$ could serve as such a paradigm of EIs. It exhibits an insulating ground state with a significant band gap, derived from an almost zero-gap normal phase. The insulating state is well reproduced by the calculations that include electron-hole Coulomb interaction. The band gap is sensitive to free carrier density, which is increased through either raising temperature or laser pumping. The gap opening is accompanied by a slight breaking of crystal symmetry with minimal lattice distortion, which has negligible effects on the band structure. These results suggest that the insulating ground state in $\rm{Ta_2Pd_3Te_5}$ originates from excitonic instability, with crystal symmetry breaking being a by-product of exciton condensation.
    
\section{Insulating ground state}\label{sec:2}
    
A previous X-ray diffraction (XRD) study reported that $\rm{Ta_2Pd_3Te_5}$ has an orthorhombic crystal structure with space group $Pnma$ (No. 62) \cite{Tremel1993}. Each unit cell consists of two $\rm{Ta_2Pd_3Te_5}$ monolayers stacked along the $a$ axis through van der Waals forces [Fig. \ref{Fig. 1}(a)]. Within each monolayer, Ta and Pd atoms form one-dimensional (1D) chains along the $b$ axis, sandwiched by Te atomic layers \cite{Tremel1993}. Consequently, the band structure from DFT calculations exhibits a quasi-1D character with strong dispersions along the chain direction [Fig. \ref{Fig. 1}(c)]. The calculations suggest a semimetallic band structure, while transport measurements reveal semiconductor-like behavior over a wide temperature range with a metal-insulator transition around 365 K [Fig. \ref{Fig. 1}(d)]. Angle-resolved photomeission specscopy (ARPES) data at 100 K in Fig. \ref{Fig. 1}(e) reveal that the valence bands (VBs) are well below $E_{\rm{F}}$, while the conduction bands (CBs) lie above $E_{\rm{F}}$, forming a global band gap. Scanning tunneling spectroscopy spectra in Fig. \ref{Fig. 1}(f) exhibit a uniform “U”-shaped gap in a clean region free from impurities. The semiconductor-like behavior with a global band gap observed in experiments contradicts the semimetallic band structure predicted by DFT calculations. 

\begin{figure*}[!t]
\centering
\includegraphics[width=0.65\textwidth]{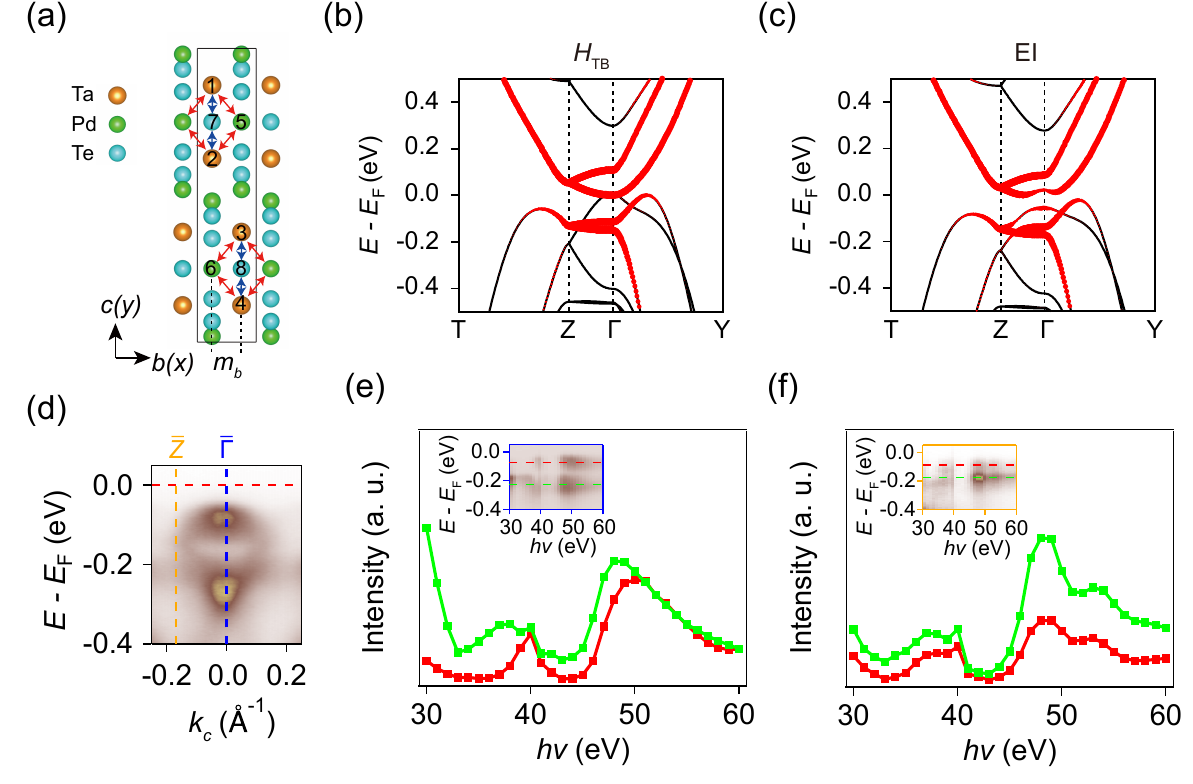}
\caption{(a) Schematic illustration of electron-hole Coulomb interaction considered in the theoretical model. Ta , Pd$_{\rm{A}}$, and Te atoms with interacting orbitals are labelled by numbers 1 $\sim$ 8. Red arrows represent the interaction between Ta-$d_{z^{2}}$ and Pd$_{\rm{A}}$-${d_{xz}}$ and blue arrows represent the interaction between Ta-$d_{z^{2}}$ and Te-$p_x$, but we assign the same interaction strength $V$ in the model. (b-c) Band structures of the noninteracting Hamiltonian $H_{\rm{TB}}$ (b) and the the EI state (c). The thickness of red lines scales the Ta-$d_{z^{2}}$ orbital component. (d) Intensity plot of the ARPES data along $\bar{\Gamma} - \bar{Z}$ collected with $h\nu$ = 40 eV at 10 K. (e) $h\nu$ dependence of the spectral intensities at two constant energies indicated by the dashed lines in the inset. Inset: Intensity plot of the ARPES data at $\bar{\Gamma}$ with varying $h\nu$. The spectral intensities are normalized by photon flux. (f) Same as (e), but collected at $\bar{Z}$.}
\label{Fig. 2}
\end{figure*}
    
In a previous study \cite{Guo2021}, Perdew-Burke-Ernzerhof (PBE) calculations suggested a semimetallic band structure with a band overlap of 85 meV. Considering the potential underestimation of the band gap by PBE, we checked the result using the modified Becke-Johnson (MBJ) functional. The MBJ band structure in Fig. \ref{Fig. 1}(c) shows a reduced band overlap of 42 meV, while still maintaining the semimetallic character (see comparison between PBE and MBJ band structures in Fig. S1 in Supplemental Material \cite{SI}). Furthermore, we confirm that the inclusion of spin-orbit coupling or changes in lattice constants is insufficient to induce a global band gap (Fig. S2 in Supplemental Material \cite{SI}). Since non-interacting DFT calculations fail to account for the band gap, correlation effects may play a crucial role in the insulating state. We conducted DFT + $U$ calculations, where $U$ represents on-site Hubbard-like electron-electron interaction on the Ta 5$d$ and Pd 4$d$ orbitals. As $U$ increases, the band overlap is moderately reduced (Fig. S3 in Supplemental Material \cite{SI}). However, even with $U$ increased to 3 eV, which is already sufficiently large for 4$d$ and 5$d$ electrons, the semimetallic band structure persists, signifying that the on-site Coulomb interaction cannot account for the insulating state. 
      
Then, we considered electron-hole Coulomb interaction for the insulating state, since the DFT calculations show a small overlap between the valence and conduction bands around $\textit{E}\rm{_F}$. For this purpose, we first constructed an effective Wannier-based two-dimensional tight-binding Hamiltonian ($H_{\rm{TB}}$) extracted from PBE calculations. The non-interacting band structure of $H_{\rm{TB}}$ is presented in Fig. \ref{Fig. 2}(b). The CB originates from the Ta-$d_{z^{2}}$ orbital, while the VB originates from the Pd$_{\rm{A}}$-${d_{xz}}$ orbital hybridized with the Te-$p_x$ orbital (Fig. S5 in Supplemental Material \cite{SI}). Here, Pd$_{\rm{A}}$ represents the Pd atoms at the A site \cite{Guo2021}. In the absence of electron-hole interaction, the hybridization between the valence and conduction bands along $\Gamma - Z$ is prohibited due to the presence of mirror symmetry $m_b$ indicated in Fig. \ref{Fig. 2}(a). Subsequently, we considered the inter-band density-density term for the electron-hole interaction $V$. The total Hamiltonian reads as follows,
\begin{equation}
\begin{aligned}
H=&H_{\rm{TB}}+VH_{int}\\
H_{int}=&\sum\limits_{\vec{R}\sigma\sigma^{\prime}}\sum\limits_{i=1,2}\left[n_{5\sigma}(\vec{R})+n_{5\sigma}(\vec{R}-\vec{b})\right]n_{i\sigma^{\prime}}(\vec{R})\\
&+\sum\limits_{\vec{R}\sigma\sigma^{\prime}}\sum\limits_{j=3,4}\left[n_{6\sigma}(\vec{R})+n_{6\sigma}(\vec{R}+\vec{b})\right]n_{j\sigma^{\prime}}(\vec{R})\\
&+\sum\limits_{\vec{R}\sigma\sigma^{\prime}}\sum\limits_{i=1,2}n_{7\sigma}(\vec{R})n_{i\sigma^{\prime}}(\vec{R})\\
&+\sum\limits_{\vec{R}\sigma\sigma^{\prime}}\sum\limits_{j=3,4}n_{8\sigma}(\vec{R})n_{j\sigma^{\prime}}(\vec{R})\\
\end{aligned}
\end{equation}
where the Ta-$d_{z^{2}}$, Pd$_{\rm{A}}$-${d_{xz}}$, and Te-$p_x$ orbitals are labelled by numbers 1 $\sim$ 4, 5 $\sim$ 6, and 7 $\sim$ 8, respectively, as indicated in Fig. \ref{Fig. 2}(a). $\sigma$ and $\sigma^{\prime}$ denote the spin degrees of freedom. We employ the Hartree-Fock approximation to treat the Coulomb interaction. In the self-consistent process, we exclude the Hartree terms, as they are already considered in DFT calculations. 
        
\begin{figure*}[!t]
\centering
\includegraphics[width=0.7\textwidth]{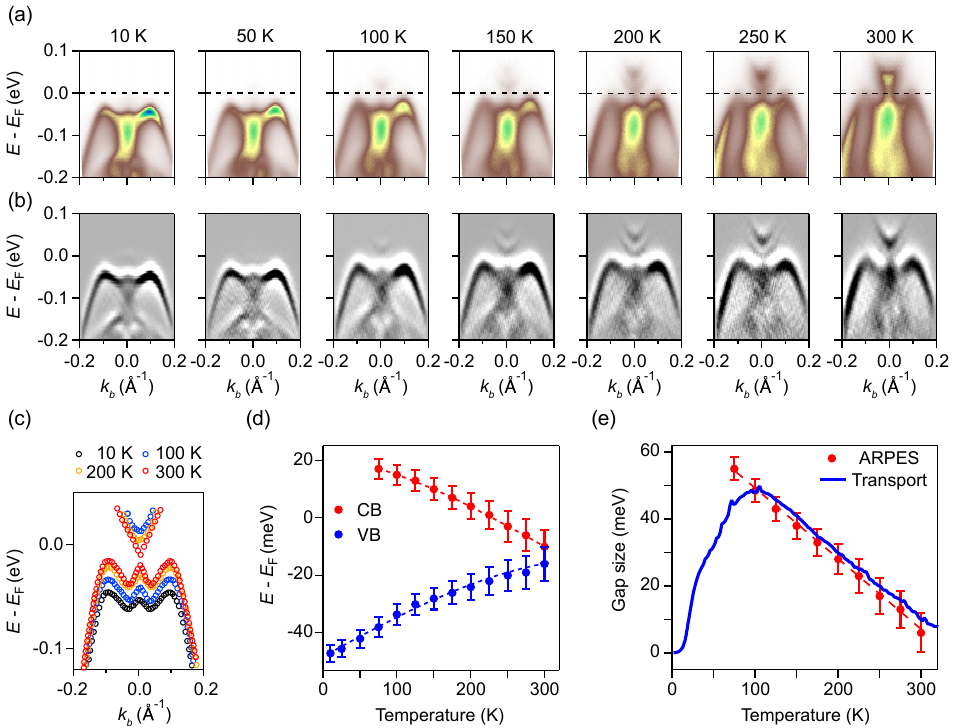}
\caption{(a) Intensity plots of the ARPES data along $\bar{\Gamma} - \bar{Y}$ collected with $h\nu$ = 6 eV at different temperatures. For clarity, the data are divided by the Fermi-Dirac distribution function to visualize the VB bottom above $E_{\rm{F}}$. (b) Intensity plots of the second derivative of the ARPES data in (a). (c) Band dispersions extracted by tracking peak positions in the intensity plots in (b). (d) Energy positions of VB top and CB bottom as a function of temperature. (e) Temperature dependence of the gap size determined from the results in (d) and $E_{\rm{A}}$ obtained by fitting the $R \sim T$ curve with the formula $E_{\mathrm{A}}=-2k_{\mathrm{B}} T^2\left(\frac{\partial \ln \rho}{\partial T}\right)$.}
\label{Fig. 3}
\end{figure*}   
    
The mean-field band structure with $V$ = 1.1 eV in Fig. \ref{Fig. 2}(c) exhibits a gap opening at $\textit{E}\rm{_F}$ with a sizable hybridization. The hybridization indicates that the $m_b$ symmetry of the electronic state is broken due to the inter-band interaction, even though crystal symmetry is preserved, which is the hallmark of excitonic instability. Here, we define an order parameter $\vec{\delta}$ of the form:
\begin{equation}
\begin{aligned}
\vec{\delta}&=(\delta_{15},\delta_{25},\delta_{36},\delta_{46})\\
\delta_{i5}
&=\langle c^{\dagger}_{i}c_{5}\rangle+\langle c^{\dagger}_{i}c_{5}(-\Vec{b})\rangle\\
\delta_{j6}
&=\langle c^{\dagger}_{j}c_{6}\rangle+\langle c^{\dagger}_{j}c_{6}(+\Vec{b})\rangle\\
\end{aligned}
\end{equation}
We have dropped the spin index as we focus on the spin-singlet case. The four $\delta_{mn}$ are in general independent. The detailed calculations show that the the inversion symmetry and all mirror symmetries are broken in the EI state, leading to a nonzero order parameter of $\vec{\delta}=\delta_0(-1,+1,-1,+1)$.

To verify the band hybridization of the EI state, we performed photon energy ($h\nu$)-dependent ARPES measurements to determine the orbital components of the VBs. For transition metal compounds, when $h\nu$ is tuned to the binding energies of \textit{p} orbitals, the spectral intensities of the VBs originating from \textit{d} orbitals are suppressed due to \textit{p}-\textit{d} antiresonance \cite{Qian2011,GaoSY2023}. The $h\nu$-dependent data at $\bar{\Gamma}$ in Fig. \ref{Fig. 2}(e) exhibit significant suppression around 34 and 43 eV, corresponding to the binding energies of the Ta 5$p_{3/2}$ and 5$p_{1/2}$ orbitals, respectively (Fig. S6 in Supplemental Material \cite{SI}). At $\bar{Z}$, in addition to the suppression around 34 and 43 eV, the data in Fig. \ref{Fig. 2}(f) show kinks around 50 and 55 eV, corresponding to the binding energies of the Pd 4$p_{3/2}$ and 4$p_{1/2}$ orbitals, respectively (Fig. S6 in Supplemental Material \cite{SI}). These results demonstrate the presence of a considerable Ta 5\textit{d} orbital component in the highest VB, signifying a strong inter-band hybridization due to the breaking of $m_b$.


\section{band gap controlled by carrier density}\label{sec:3}   

\begin{figure*}[!t]
\centering
\includegraphics[width=0.85\textwidth]{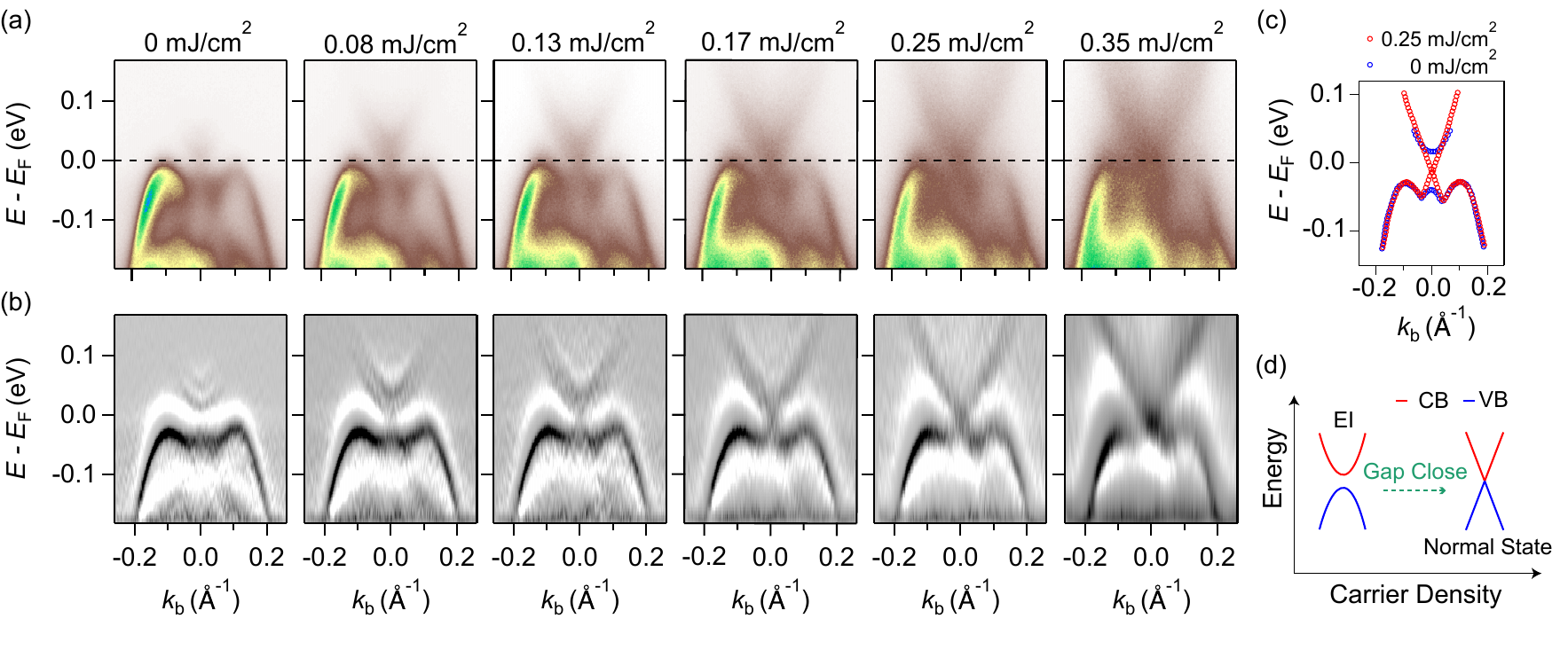}
\caption{(a) Intensity plots of the ARPES data along $\bar{\Gamma} - \bar{Y}$ collected with different pump fluences. All non-equilibrium data were collected at a time delay of 0 ps. (b) Intensity plots of the second derivative of the ARPES data in (a). (c) Band dispersions extracted by tracking peak positions in the intensity plots in (b) at pump fluences of 0 and 0.25 $\rm{mJ/cm^2}$. (d) Schematic diagram depicting the transition from an EI to normal state with increasing carrier density.}
\label{Fig. 4}
\end{figure*}
    
EIs exhibit a semiconductor-like band structure with a band gap at $E_{\rm{F}}$, where thermal excitation leads to an increase in the number of free carriers with increasing temperature. The higher carrier density effectively screens the electron-hole Coulomb interaction, diminishing the stability of excitons. As a result, the gap size of an EI gradually decreases as temperature increases \cite{Cercellier2007,Monney2010,Seki2014}. Hall resistance measurements on $\rm{Ta_2Pd_3Te_5}$ reveal a substantial increase in carrier density with increasing temperature above 30 K (Fig. S10 in Supplemental Material \cite{SI}). We performed temperature-dependent ARPES experiments to investigate the evolution of the band structure with temperature. The ARPES data in Figs. \ref{Fig. 3}(a) and \ref{Fig. 3}(b) show that the VBs lie well below $E_{\rm{F}}$ at low temperatures, while the CBs above $E_{\rm{F}}$ become visible when thermally occupied above 100 K. Both valence and conduction bands shift in energy with temperature [Fig. \ref{Fig. 3}(c)]. They gradually approach with increasing temperature [Fig. \ref{Fig. 3}(d)], resulting in a substantial reduction in the gap size from 55 meV at 75 K to 5 meV at 300 K [Fig. \ref{Fig. 3}(e)]. Furthermore, the dispersions around $\bar{\Gamma}$ change from parabolic to linear when the temperature reaches 300 K. These results reveal a significant renormalization in the low-energy state with temperature.
    
The temperature dependence of the gap size is further validated by transport data. The $R \sim T$ curve in Fig. \ref{Fig. 1}(d) exhibits semiconductor-like behavior below 365 K. In semiconductors, the thermal activation of carriers follows the relation $\rho = \rho_0 \exp \left(-\frac{E_{\mathrm{A}}}{2k_{\mathrm{B}} T}\right)$, where $\rho_0$ is a constant with the dimension of resistance, and $E_{\rm{A}}$ represents the activation energy of carriers, reflecting the magnitude of the band gap. By fitting the $R \sim T$ curve with the formula $E_{\mathrm{A}}=-2k_{\mathrm{B}} T^2\left(\frac{\partial \ln \rho}{\partial T}\right)$, we obtain $E_{\rm{A}}$ as a function of temperature [Fig. \ref{Fig. 3}(e)]. Above 100 K, $E_{\rm{A}}$ monotonically decreases with increasing temperature, and its magnitude closely aligns with the gap size extracted from the ARPES data, indicating that the thermal activation model effectively describes the evolution of the band gap above 100 K. 
    
In contrast, $E_{\rm{A}}$ decreases dramatically below 100 K. This behavior can be attributed to the presence of in-gap states, which contribute finite intensities near $E_{\rm{F}}$ to the low-temperature ARPES spectra (Fig. S9 in Supplemental Material \cite{SI}). The in-gap states could originate from impurities with a finite spatial extension of the wave function, as observed in the Si-doped semiconductor $\beta$-$\rm{Ga_2O_3}$ \cite{Richard2012,Iwaya2011}. Based on this assumption, we estimate the spatial extension of the impurity wave function to be about 16 \AA\ by fitting the momentum distribution curve at $E_{\rm{F}}$ (Fig. S11 in Supplemental Material \cite{SI}). Scanning tunneling microscopy (STM) measurements confirm the presence of impurities in some regions on sample surfaces, and their spatial extensions are consistent with the value estimated from the ARPES data (Fig. S12 in Supplemental Material \cite{SI}). Carrier hopping between nearby impurities would have a significant impact on transport behavior at low temperatures, leading to an obvious deviation from the thermal activation model as well as a rapid upturn in carrier density below 30 K (Fig. S10 in Supplemental Material \cite{SI}. With increasing temperature, the contribution of thermally excited intrinsic carriers gradually rises, eventually becoming dominant above 100 K.

\begin{figure*}[!t]
\centering
\includegraphics[width=0.75\textwidth]{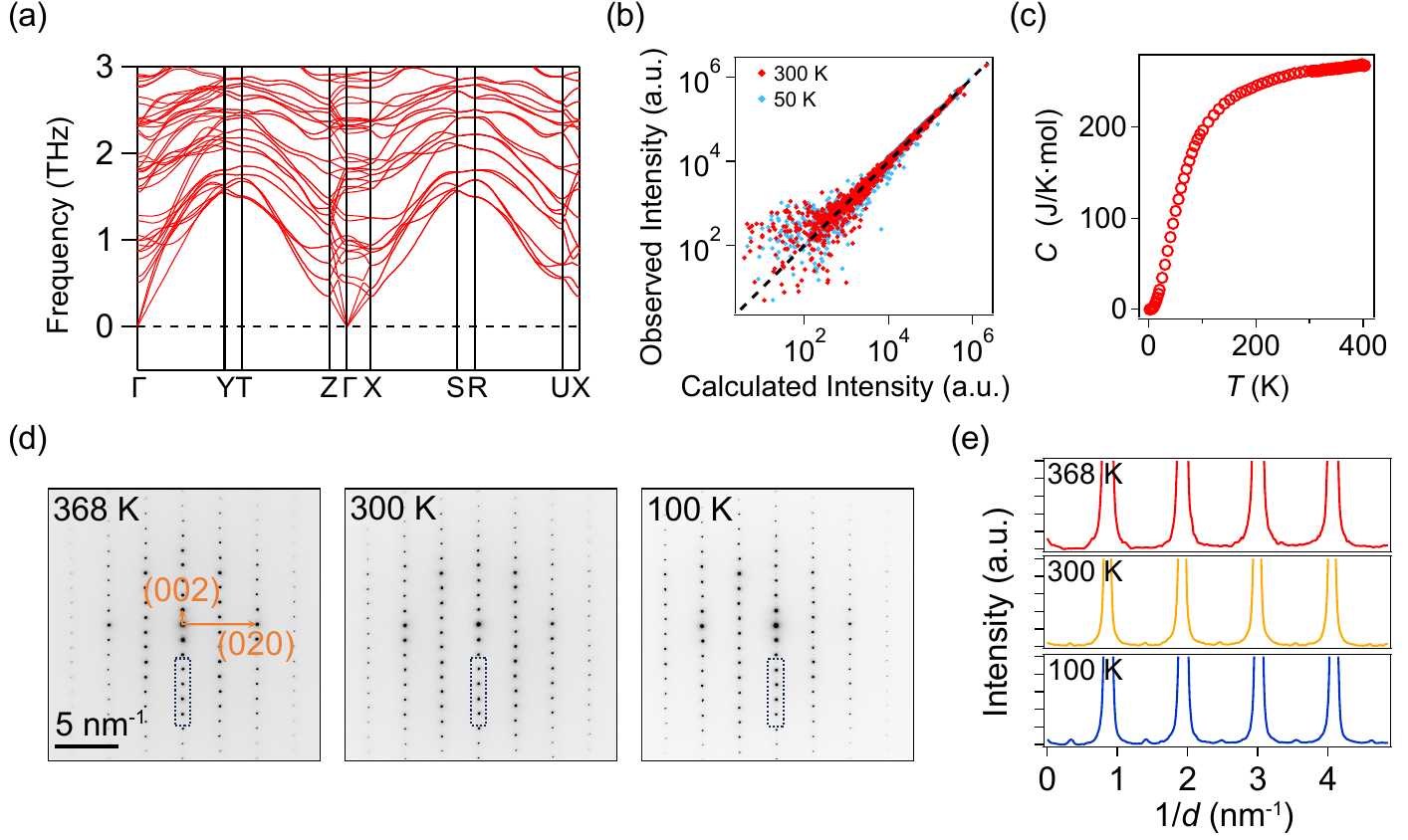}
\caption{(a) Calculated phonon dispersions of $\rm{Ta_2Pd_3Te_5}$ along high-symmetry lines by DFT using the GGA functional. (b) Comparison between the observed XRD intensities from all Bragg peaks measured at 300 and 50 K and the calculated intensities with space group $Pnma$. (c) Specific heat as a function of temperature. (d) Electron diffraction patterns oriented with respect to the [100] direction at 368, 300, and 100 K. (e) Line profiles along the [001] direction extracted from the regions in the rectangles in (d).}
\label{Fig. 5}
\end{figure*} 
    
In addition to heating in the equilibrium state, in pump-probe ARPES experiments, pump laser pulses can substantially raise the transient electronic temperature \cite{Tang2020,Okazaki2018}, thereby increasing the number of thermally excited carriers in the non-equilibrium state (Fig. S14 in Supplemental Material \cite{SI}). Hence, we investigate the effect of laser pumping on the insulating state of $\rm{Ta_2Pd_3Te_5}$ with pump-probe ARPES. To observe the evolution of the CBs more clearly and reduce the thermal effect (Fig. S13 in Supplemental Material \cite{SI}), the pump-probe ARPES experiments were conducted at 150 K. The ARPES data in Figs. \ref{Fig. 4}(a) and \ref{Fig. 4}(b) exhibit a clear band gap in the equilibrium state without laser pumping. The gap gradually diminishes with an increase in pump fluence and is closed at a fluence of 0.17 $\rm{mJ/cm^2}$. We compare band dispersions at pump fluences of 0 and 0.25 $\rm{mJ/cm^2}$ in Fig. \ref{Fig. 4}(c). The dispersions around $\bar{\Gamma}$ change from curved to linear, resulting in a gap closing, while those away from $\bar{\Gamma}$ are almost unchanged. 

Our ARPES results have demonstrated that the band gap is closed with a significant change in the dispersions around $\bar{\Gamma}$, when the number of carriers becomes sufficiently large through thermal excitation in both equilibrium and non-equilibrium states, as illustrated in Fig. \ref{Fig. 4}(d). These findings are well consistent with the expectation for EIs.   

\section{Structural Instability}\label{sec:4}

In an EI state, structural instability is expected to occur due to finite electron-lattice coupling. Hence, we conducted a systematic investigation to clarify whether structural instability exists in $\rm{Ta_2Pd_3Te_5}$. The calculated phonon spectra in Fig. \ref{Fig. 5}(a) reveal the absence of imaginary frequencies, indicating that the $Pnma$ crystal structure is stable with no tendency towards spontaneous symmetry breaking. This result is in contrast to the calculations for the EI candidates 1\textit{T}-$\rm{TiSe_2}$ \cite{Calandra2011,Duong2015,Otto2021} and $\rm{Ta_2NiSe_5}$ \cite{Subedi2020,Baldini2020,Windgatter2021}. The single crystal XRD data at 300 and 50 K can be well refined with space group $Pnma$ [Fig. \ref{Fig. 5}(b)], in agreement with the previous report \cite{Tremel1993}. The Raman spectra exhibit continuous temperature-dependent behavior for all phonon modes without any splitting or emergence of new modes below 450 K (Fig. S15 in Supplemental Material \cite{SI}). The specific heat data in Fig. \ref{Fig. 5}(c) do not show any anomaly associated with a structure phase transition. These results point to a stable crystal structure without a structural phase transition. We note that an electronically driven EI phase transition cannot be ruled out in the specific heat data, since the contribution of electronic heat capacity is negligible at high temperatures. 

While no structural phase transition is identified, transmission electron microscopy (TEM) measurements reveal the existence of crystal symmetry breaking in the insulating state. For space group $Pnma$, the (\textit{h}00) and (00\textit{l}) diffraction spots, where \textit{h} and \textit{l} are odd numbers, are forbidden due to the presence of glide symmetries $\bar{m}_c$ and $\bar{m}_a$. The electron diffraction patterns oriented along the [100] direction are displayed in Fig. \ref{Fig. 5}(d). To demonstrate the subtle changes with temperature in these patterns, we highlight the base of the line profiles along the [001] direction in Fig. \ref{Fig. 5}(e). Only the primary spots with an even number of \textit{l} are observed at 368 K, confirming that $\bar{m}_a$ is preserved. Remarkably, the (00$l$) forbidden spots become visible at 300 and 100 K. Similar behavior is observed for the (\textit{h}00) forbidden spots (Fig. S16 in Supplemental Material \cite{SI}). The emergence of the (\textit{h}00) and (00\textit{l}) forbidden spots upon cooling indicating that both $\bar{m}_c$ and $\bar{m}_a$ are broken. The simulated electron diffraction patterns for $Pnma$ and its orthorhombic subgroups indicate that both (\textit{h}00) and (00\textit{l}) forbidden spots appears only in space group $P2_12_12_1$ (Fig. S17 in Supplemental Material \cite{SI}), in which both $\bar{m}_c$ and $\bar{m}_a$ are broken.

The crystal symmetry breaking occurs near the metal-insulator transition with a gap opening, signifying a close relationship between electronic and structural instabilities in $\rm{Ta_2Pd_3Te_5}$. In the EI candidates 1\textit{T}-$\rm{TiSe_2}$ and $\rm{Ta_2NiSe_5}$, the concurrence of electronic and structural instabilities has aroused intense debate on the origin of the gap opening \cite{Subedi2020,Baldini2020,Hughes1977,Motizuki1986,Rossnagel2002,Watson2020,Kim2020,Chen2022}. While crystal symmetry breaking allows the opening of a hybridized gap between valence and conduction bands, the gap size depends on the magnitude of lattice distortion. By comparing the refinement results of the XRD data at 50 K with $Pnma$ and $P2_12_12_1$, we confirm that the lattice distortion is exceptionally small in the symmetry-breaking phase (Tables S1$\rm{-}$S3 in Supplemental Material \cite{SI}). To assess the impact of crystal symmetry breaking on the band structure, we calculated the band structure of $P2_12_12_1$ with the refined lattice parameters. The band structure remains almost unchanged due to the minimal lattice distortion (Fig. S18 in Supplemental Material \cite{SI}). Therefore, the slight breaking of crystal symmetry cannot account for the significant gap opening. Given the stable crystal structure and the negligible effects of crystal symmetry breaking on the band structure, we rule out the possibility that the gap opening in $\rm{Ta_2Pd_3Te_5}$ is induced by a structural change resulting from phononic instability.

In contrast, the gap opening with crystal symmetry breaking can be well explained within the framework of EIs. The high-temperature normal phase exhibits an almost zero-gap band structure with linear dispersions. The electron-hole Coulomb interaction is weakly screened due to the quasi-1D structure with a vanishing density of states at $E_{\rm{F}}$. The electronic state becomes unstable against exciton condensation. Excitons are trapped to lattice sites and form a static charge transfer from the Pd 4$d$/Te 5$p$ to Ta 5$d$ orbital, breaking the symmetry of the electronic state. This allows the hybridization between the valence and conduction bands, resulting in the opening of a many-body gap. Meanwhile, the electronic symmetry breaking induces local lattice distortion through electron-lattice coupling, causing a breaking of crystal symmetry. The small lattice distortion suggests a weak coupling between electron and lattice degrees of freedom. 

\section{Outlook}\label{sec:5}

We have revealed that $\rm{Ta_2Pd_3Te_5}$ exhibits an EI ground state emerging from an almost zero-gap normal phase. As the VB top and CB bottom are located at the same momentum point $\Gamma$ in the BZ (Fig. S4 in Supplemental Material \cite{SI}, the excitons carry zero momentum. The zero-momentum excitons couple to the $\boldsymbol{q}$ = 0 phonon mode, resulting in crystal symmetry breaking that does not change the period in the lattice. It is well known that BCS-type exciton condensation occurs in a semimetal, while BEC-type  exciton condensation occurs in a semiconductor. Our results show that the normal phase features an almost zero-gap band structure, making $\rm{Ta_2Pd_3Te_5}$ a promising platform for investigating many-body phenomena in the BCS-BEC crossover region. Due to the quasi-1D van der Waals structure, this material can be easily mechanically exfoliated into large flakes with long straight edges along the 1D chains \cite{Wang2022}. The chemical potential can be conveniently tuned by applying gate voltages to thin flakes, which is favorable for further studies of the EI state. Consistently, the first principles calculations with many-body perturbation theory predict an EI state in the $\rm{Ta_2Pd_3Te_5}$ monolayer \cite{Yao2024}. Furthermore, the abnormal band order indicates that $\rm{Ta_2Pd_3Te_5}$ is an unconventional material with a mismatch between charge centers and atomic positions \cite{Gao2022,Guo2022}, leading to the emergence of quasi-1D edge states with typical Luttinger liquid behavior within the excitonic gap \cite{Wang2022}.

\section{Method}\label{sec:5}

\subsection{Sample growth}

Single crystals of $\rm{(Ta_2Pd_3Te_5}$ were synthesized by the self-flux method. Starting materials of Ta (powder, 99.999\%), Pd (bar, 99.9999\%) and Te (lump, 99.9999\%) were mixed in an Ar-filled glove box at a molar radio of Ta : Pd : Te = 2 : 4.5 : 7.5. The mixture was placed in an alumina crucible, which was then sealed in an evacuated quartz tube. The tube was heated to 950 °C over 10 hours and dwelt for 2 days. Then, the tube was slowly cooled down to 800 °C at a rate of 0.5 °C/hour. Finally, the extra flux was removed by centrifuging at 800 °C. After centrifuging, the black and shiny single crystals of $\rm{(Ta_2Pd_3Te_5}$ were picked out from the remnants in the crucible.
    
\subsection{ARPES experiments}        
Laser-based ARPES experiments were conducted at the Institute of Physics, Chinese Academy of Sciences \cite{Pan2023}. The equilibrium ARPES data were collected with $h\nu$ = 6 eV. In non-equilibrium ARPES experiments, the samples were pumped by an ultrafast laser pulse ($h\nu$ = 2.4 eV) with a pulse duration of 250 fs and a repetition rate of 800 kHz. An ultraviolet probe laser pulse ($h\nu$ = 7.2 eV) subsequently photoemits electrons. The overall time and energy resolutions were set to 500 fs and 18 meV, respectively. Synchrotron ARPES experiments were conducted at the 03U beam line \cite{Yang03U} and the “Dreamline” beam line at the Shanghai Synchrotron Radiation Facility.

\subsection{STM experiments}
STM experiments were conducted in a low temperature ultra-high vacuum STM system, Unisoku USM-1300. Topographic images were acquired in the constant-current mode with a tungsten tip. Before the measurements, STM tips were heated by e-beam and calibrated on a clean Ag surface. The scanning settings of all the topographies were under a bias voltage of 100 mV and a setpoint of 100 pA unless specifically mentioned. Differential conductance spectra were acquired by a standard lock-in technique at a reference frequency of 973 Hz unless specifically mentioned.

\subsection{TEM experiments}
The specimens oriented with respect to the [100] direction were thinned through conventional mechanical exfoliation method, while those oriented with respect to the [001] direction were prepared by conventional focused ion beam (FIB, FEI Helios 600i) lift-out method. After that, the specimens were transferred to a Cu grid for ED characterization, which were collected using a transmission electron microscope (TEM, JEOL JEM-F200). The cryo-TEM holder (Gatan) with heating controller was used to maintain the specimen temperature at 100, 300 and 368 K. The diffraction patterns at each temperature were collected using an operating voltage of 200 kV at a camera length of 400 mm.

\subsection{X-ray diffraction experiments}
A specimen of $\rm{Ta_2Pd_3Te_5}$ with dimensions of 0.017 mm $\times$ 0.038 mm $\times$ 0.089 mm was used for the X-ray crystallographic analysis on the Bruker D8 Venture with the Mo K$\alpha$ radiation ($\lambda$ = 0.71073 \AA. The frames were integrated with the Bruker SAINT software package using a narrow-frame algorithm. The data were corrected for absorption effects using the MultiScan method (SADABS). Data collection, cell refinement, and data reduction were performed using the Bruker APEX4 program. The refinement was carried out using SHELX programs within the Olex2-1.5-alpha software package \cite{Dolomanov2009}.  The crystal structure was successfully solved using the intrinsic phasing method with SHELXT \cite{Sheldrick2015_1} and refined with SHELXL \cite{Sheldrick2015_2} against the $\rm{F^2}$ data, incorporating anistropic displacement parameters for all atoms.
    
\subsection{DFT calculations} 
First-principles calculations were conducted within the framework of DFT using the projector augmented wave (PAW) method \cite{Blochl1994,Kresse1999}, as implemented in Vienna $\textit{ab initio}$ simulation package (VASP) \cite{Kresse1996,Kresse1996_2}. The PBE generalized gradient approximation exchange-correlations functional \cite{Perdew1996} was used. In the self-consistent process, 4 $\times$ 16 $\times$ 4 $k$-point sampling grids were used, and the cut-off energy for plane wave expansion was 500 eV. Since PBE band calculations usually underestimate the band gap, we introduce the MBJ exchange potential \cite{Becke2006,Tran2009} to improve this underestimation. The maximally localized Wannier functions (MLWFs) were constructed by using Wannier90 package \cite{Pizzi2020}.

\section*{Acknowledgements}
We thank Xi Dai, Xinzheng Li, Hongming Weng, Xuetao Zhu and Peng Zhang for useful discussions. We thank Zhicheng Rao, Anqi Wang, Huan Wang, Xuezhi Chen, Bingjie Chen, Xiaofan Shi, Xingchen Guo, Zhe Zheng, Mingzhe Hu, Yao Meng, Hongxiong Liu, Xiafan Xu, Jin Ding for technical assistance. This work was supported by the Ministry of Science and Technology of China (Grants No. 2022YFA1403800, No. 2022YFA1402704 and No. 2022YFA1403400), the National Natural Science Foundation of China (Grants No. U1832202, No. U22A6005, No. U2032204, No. 12188101, No. 92065203, No. 12174430, No. 12274186, No. 11888101, No. 12104491, No. 11774419, and No. 52272268), the Chinese Academy of Sciences (Grants No. XDB33000000, No. XDB28000000 and No. GJTD-2020-01), the Beijing Natural Science Foundation (Grant No. JQ23022), the Beijing Nova Program (Grant No. Z211100002121144), the New Cornerstone Science Foundation (Grant No. 23H010801236), the Innovation Program for Quantum Science and Technology (Grant No. 2021ZD0302700), the Shanghai Committee of Science and Technology (Grant No. 23JC1403300), the China Postdoctoral Science Foundation funded project (Grant No. 2021M703461), the Shanghai Municipal Science and Technology Major Project, the Synergetic Extreme Condition User Facility, and the Centre for Materials Genome. Q. M. Z. acknowledges the support from Users with Excellence Program of Hefei Science Centre and High Magnetic Field Facility, CAS. S. Y. G. was supported by the China Scholarship Council (Grant No. 202104910090) and the Sino Swiss Science and Technology Cooperation (Grant No. CN-EG-03-012021).

\end{document}